\documentclass{article}
\usepackage[utf8]{inputenc}
\usepackage{geometry}
\usepackage{setspace}
\usepackage{titlesec}
\usepackage{natbib}
\usepackage{authblk}
\setcitestyle{authoryear, open={((},close={)}}
\usepackage{threeparttable}
\usepackage{graphicx}
\usepackage{gensymb}
\usepackage{xcolor, soul}
\usepackage{pdfpages}
\usepackage{colortbl}
\usepackage{nicematrix}
\usepackage{longtable}
\usepackage{ifthen}
\usepackage{lineno}
\usepackage[singlelinecheck=false]{caption}

\title{Assessing global drivers of forest transpiration using clustered machine learning models}

\author[a]{Morgan Thornwell}
\author[a]{David Yang}
\author[a]{Cheng-Wei Huang}
\author[a]{Peyman Abbaszadeh}
\author[a,*]{Samantha Hartzell}

\affil[a]{Department of Civil and Environmental Engineering, Portland State University, Portland Oregon 97201, USA}

\affil[*]{Corresponding author: Samantha Hartzell, email: s.hartzell@pdx.edu;
postal address: Suite 200, 1930 SW 4th Ave, Portland OR 97210; phone: (503) 725-2435}

\begin{document}

\pagenumbering{gobble}

\maketitle

\clearpage


\section*{Abstract}

Understanding the environmental drivers of forest transpiration is critical for improving global predictions of water availability and ecosystem health. Due to the many competing controls on plant water stress and ecosystem transpiration, however, these drivers may vary widely across tree species which have adapted hydraulically to local climate conditions. Here, clustered machine learning models were used to analyze global drivers of forest transpiration rates using the global SAPFLUXNET database. Sap flux data from a total of ninety-five sites spanning seven biomes were grouped using two clustering strategies: by biome and by plant functional type. Two supervised machine learning algorithms, a random forest algorithm and a neural network algorithm, were used to predict rates of sap flux for each cluster. The performance and feature importance in each model were analyzed and compared to evaluate the environmental variables that control each cluster's performance. By defining site clusters, these machine learning models are able predict transpiration across a wide variety of geographical sites and tree species and the environmental variables that control it. Unlike machine learning models trained on the entire dataset, high-performing cluster-based models achieved $R^2$ values to measurement data in the range of 0.74 to 0.90, with the highest performance being achieved in mid-sized clusters of up to thirty-five to thirty-six sites. While most clusters showed good performance, there was high variance in feature importance between clusters, indicating that key predictors of transpiration varied strongly across both plant functional type and biome. Overall, water-limited climates tended to be more controlled by soil water content, whereas climates with high mean annual temperature tended to be more controlled by solar radiation and less dependent on air temperature. These findings provide insights into how forest transpiration responds to environmental factors across a wide range of climate types and tree species.
\\

\noindent
Keywords: Transpiration, SAPFLUXNET, random forest, neural network, clustering, plant functional type, biome

\clearpage


\setcounter{page}{1} 
\pagenumbering{arabic}
\section{Introduction}

Forest evapotranspiration (ET) is typically considered to be a large portion of the hydrological budget, although estimates of forest ET as a fraction of precipitation vary widely, ranging from less than 20\% to more than 90\% across various sites \citep{komatsu2012simple}. Accurate measurement of ET and its component fluxes can be difficult and expensive, and is often undertaken at large scales through eddy covariance measurements, which provide a measure of total ET with relatively large uncertainty \citep{Wilson2001}. While it is widely understood that the controls of vegetation on ET heavily impact the hydrologic cycle and influence the regional and global climate \citep{Bonan2008}, partitioning of ET into its component fluxes of transpiration and evaporation also introduces a large degree of uncertainty \citep{berg2019evapotranspiration}, complicating our understanding of and ability to predict this process. An improved understanding of forest transpiration and its environmental and biophysical controls is required for better partitioning of ET and improved predictions of vegetation water fluxes, which can then be leveraged to aid in managing water resources \citep{casagrande2018assessing}, predicting carbon uptake \citep{krause2018large}, improving energy partitioning \citep{Bonan2008}, anticipating the effects of climate change \citep{alkama2016biophysical}, and improving remote sensing ET data products \citep{Chen2020evolution} among other endeavors. 

The use of Machine Learning (ML) models for predicting transpiration has been studied in recent years as a way to reduce data requirements, make more accurate predictions, and understand which environmental variables exert primary controls on transpiration. ML models that have been used for this purpose include support vector machines \citep{Duo2018, Ferreira2019, Granata2019, Fan2021}, artificial neural networks \citep{Liu2009, Xu2017, Ellsaber2020}, random forests \citep{Granata2019, Ellsaber2020, Shao2021}, and gene expression programming \citep{Kiafar2017, Landeras2018}. These ML models not only serve as effective tools to predict transpiration, but also provide insights into the response of transpiration to environmental variables across different biomes and plant functional types (PFTs). 

Many of the studies that implement ML algorithms to estimate transpiration use analytical models, most notably the FAO-56 Penman-Monteith (PM) method, to generate reference values of transpiration (i.e., target values in ML). Given enough data generated from the PM model, a wide variety of ML algorithms are capable of predicting the target values based on a variety of environmental input variables. Yet, the added value of ML models is not well justified in such cases because these ML models at their best simply reproduce the results of the PM model (not true field data). Specifically, ML algorithms will only recover the prescribed form of the PM model based on \textit{a priori} assumptions regarding how the involved variables (e.g., stomatal conductance) respond to environmental factors. However, the prediction of stomatal conductance remains a vexing problem \citep{buckley2017modeling}. As a key regulator of transpiration, stomatal conductance is impacted not only by environmental factors, but also by plant attributes, such as plant water potential, xylem conductance and vulnerability, and stomatal regulation strategy \citep{sperry2015plant, liu2020plant}. It is unclear whether the target values obtained from the PM model correspond well to actual transpiration because careful calibration, measurement of water stress, and crop coefficients to describe stomatal conductance are often missing in such studies. For instance, a comparison between a random forest model, trained on the FAO-56 reference ET for maize, to field ET data obtained using a soil water balance method and found only a weak correlation between the ML predictions and the field data ($R^2$ between 0.48 and 0.77) \citep{Shao2021}. Thus, direct measurements of transpiration (to be used as target values) are required when using ML algorithms to predict transpiration and subsequently assess how transpiration is impacted by various environmental variables across different biome and PFTs in forest ecosystems. 

Studies have increasingly opted to address this problem by using direct field measurements, including lysimeter measurements \citep{Kumar2002}, eddy covariance data \citep{Duo2018, Chen2020}, and sap flux data \citep{Liu2009, Tu2019, Ellsaber2020} as target values for training. The recently published SAPFLUXNET database, a global collection of sap flux measurements, environmental data, and metadata \citep{Poyatos2020}, has enabled further advances in modeling and prediction of transpiration, including the development of Long Short-Term Memory networks generalized to the European subcontinent \citep{Loritz2024}, hybrid models incorporating SAPFLUXNET data with eddy covariance and satellite observations \citep{Koppa2022}, evaluation of existing transpiration products \citep{Bittencourt2023, Li2024}, and analysis of environmental drivers of canopy conductance \citep{flo2022vapour}. This database provides direct measurements of transpiration (rather than total ET) from deciduous, mixed, and evergreen forests. With this database, containing 174 species and a wide range of climatic conditions, there is an opportunity to fill in the aforementioned research gap regarding ML models trained on field measurements. This study builds on such work by leveraging SAPFLUXNET data to develop a suite of ML models that estimate transpiration rates across diverse species and climates. 

It is expected that the use of the SAPFLUXNET database may shed light on the ability of ML algorithms to capture trends across various locations and species due to its characterization of a large number of species, geographical locations, and climate types. A common challenge for ML models is that, unlike common mechanistic models, they typically do not generalize well to multiple climates or locations that are drastically different from the study regions. For instance, \cite{Karimi2017} was only able to generalize ML models to other locations with a similar, humid climate. Researchers have attempted to solve this challenge by incorporating data from multiple locations into one single ML model \citep{Yan2021}. Another approach involves implementing a k-means clustering algorithm to create climate clusters, and then training separate models for each cluster \citep{Ferreira2019}. Most studies taking these approaches have used target values from the PM model, which does not necessarily reflect actual transpiration rates, rather than true field data. These studies did indicate that the ML models, when trained based on diverse types of climates, are capable of adequate generalization. Yet, similar endeavors have not been undertaken to study how well ML models that were trained on true field data across many forested ecosystems may shed light on climate drivers of transpiration.

In this study, we select two ML algorithms, artificial neural networks and random forest, to predict transpiration using the SAPFLUXNET database. Artificial neural networks, particularly deep neural networks, are by far the most popular choice for transpiration modeling due to their ability to model complex, nonlinear relationships \citep{Shen2018}. Random forest models are also widely used in the literature because they are easy to implement \citep{jeong2016random}. We compare results predicted by training multiple deep neural network and random forest models on the SAPFLUXNET database, with target data clustered based on different biome and PFTs. After creating models for each cluster, we also perform an in-depth investigation into model accuracy and feature importance in predicting transpiration rates. This analysis elucidates which variables control transpiration across varying locations and PFTs, and informs appropriate data clustering strategies for future work in forest transpiration modeling. 

\section{Materials and Methods}

The workflow for building and training ML models was to select relevant input features from the available data, separate data by geographic site using biome and species type attributes, select ML algorithms to build the models, and then calculate the metrics that can be used to describe the models' performance and sensitivity. A separate ML model was built for each group of sites, and the performance metrics and feature importance for each model were compared. The selected ML algorithms were neural networks and random forests. The $R^2$ value of the predicted transpiration rates to the measured transpiration rates was used as the primary indicator of model performance. Feature importance tests were used to describe each model's sensitivity to the input features.

\subsection{Input features and target data}
     
\begin{figure}
    \centering
    \includegraphics[width=16cm]{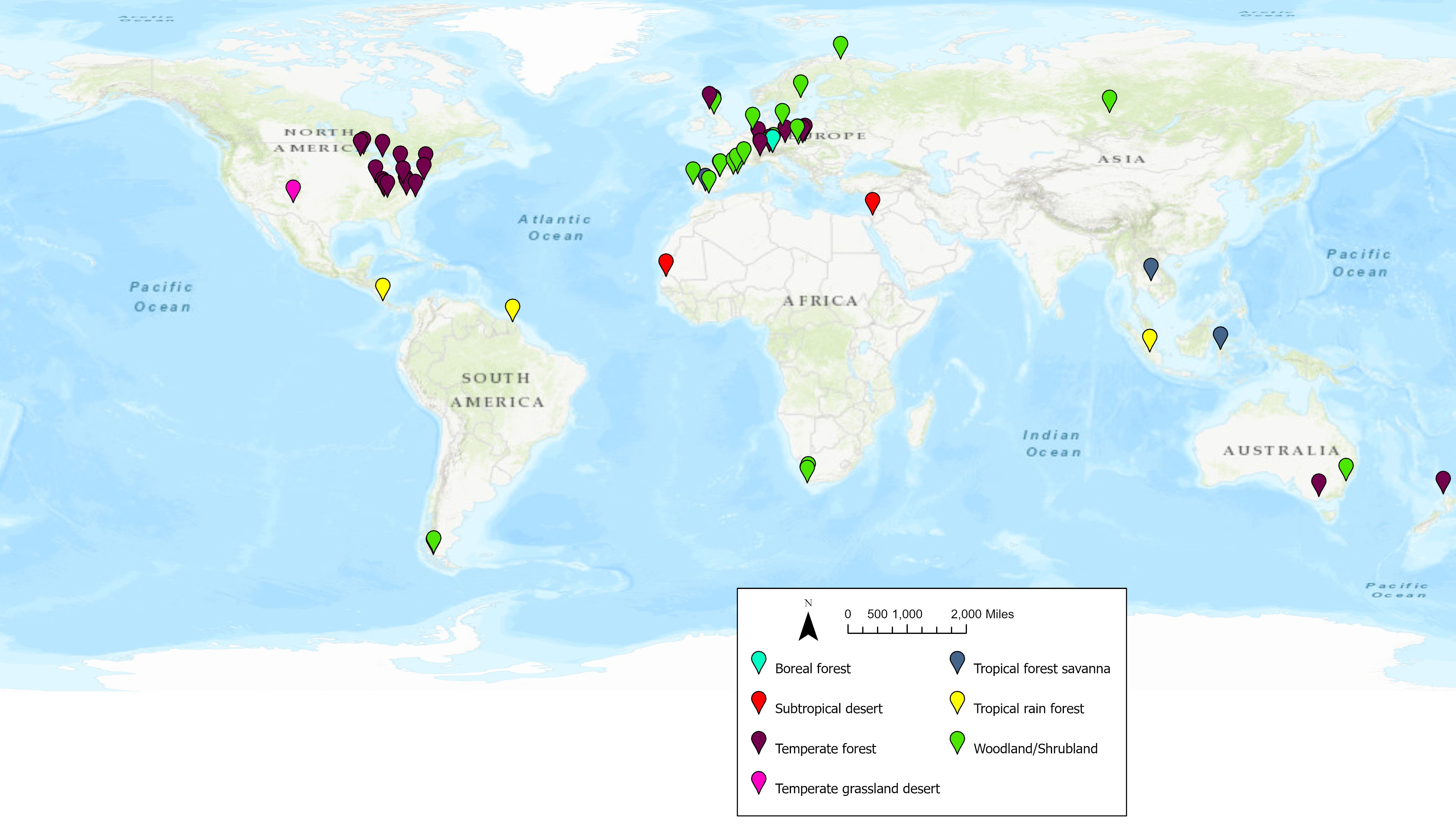}
    \caption{Geographic distribution of represented data by biome type, representing 95 sites with complete sets of environmental variables.}
    \label{fig:site distribution}
\end{figure}

The environmental variables considered for use as input features were air temperature in \degree C, vapor pressure deficit (VPD) in kPa, relative humidity (RH) in kPa kPa$^{-1}$, shallow soil water content (SWC) in m$^{3}$ m$^{-3}$, precipitation in mm and photosynthetic photon flux density (PPFD) in $\mu$mol s$^{-1}$m$^{-2}$. Each of these environmental variables were obtained from the SAPFLUXNET database at a half-hourly interval. Sites that did not have the complete set of environmental variables have been removed from consideration. As a result, the number of sites considered is 95 (Fig. \ref{fig:site distribution}). Locations in the Eastern U.S. and Western Europe, which tend to have temperate forest or woodland/shrubland biomes, are over-represented in the database, yet the site distribution covers a wide range of environmental conditions, including data on every continent except Antarctica. Seven unique biomes, including woodland/shrubland, temperate forest, tropical rainforest, tropical forest savanna, subtropical desert, temperate grassland desert, and boreal forest, are represented in the data.

Due to the strong interdependence among air temperature, RH, and VPD \citep{monteith2013principles}, co-linearity may occur between these variables. The preliminary analysis indicated little difference in model performance whether removing RH or removing VPD from the set of input features. Thus, RH was removed from the list of input features, considering that VPD is a primary driving force for transpirational flow \citep{liu2020plant, flo2022vapour}. Similarly, precipitation had a negligible feature importance (0.002-0.014 for neural networks and 0.002-0.007 for random forests) to the prediction accuracy of all trained ML models when soil moisture was included, suggesting that soil moisture is a better representation of water supply for transpiration. For this reason, precipitation was also not considered as one of the input features. The final list of input features selected for use in the ML algorithms included air temperature, VPD, SWC, and PPFD. Their values are available with a half-hour interval during the period of measurement.

The target value used to train the ML algorithms was transpiration rate, which was inferred from the whole-plant sap flow collected from the SAPFLUXNET database. The whole-plant scaled data were determined from raw measurements of sap flow from a variety of heat-based sensors and sap flow models combined with whole-plant scaling \citep{Poyatos2020}. The sap flows (cm$^3$ h$^{-1}$) of all trees at a site were averaged for each half-hourly time step and used as the target value for that site during the development of ML models. 

\subsection{Clustering strategy}

The sites were grouped into clusters and the clustering strategy was implemented using two approaches: grouping by PFT and by biome. The full list of sites with their respective classifiers by biome and functional type is included in Supplementary Information A. To test which clustering strategy would yield the most accurate predictive model and measure differences in sensitivity between clusters, each of the clusters was associated with two different ML models.

Within the SAPFLUXNET database, there were 95 species. At some sites, only one species was present (some of which were also present at other sites) and at others there were multiple species present. Therefore, instead of directly grouping by species, PFTs were used so that the clusters would contain sufficient data for a ML algorithm to be effective and potentially generalize to new sites. Based on the functional type recorded in the SAPFLUXNET, 42$\%$ of the database is evergreen, 37$\%$ is deciduous, 19$\%$ is mixed species, and 2$\%$ is not classified. Among the 95 sites considered here, there are 37 unique species present, with an almost equal distribution among different PFTs (see Supplementary Information A). 

Alternatively, seven different biome types were used to group data separately by climate conditions. 38$\%$ of the data used is classified as woodland/shrubland, 41$\%$ as temperate forest, 5$\%$ as tropical rainforest, 5$\%$ as tropical forest savanna, 4$\%$ as subtropical desert, 3$\%$ as boreal forest, and 3$\%$ as temperate grassland desert. Some biomes were better represented than others as a result of data availability in the SAPFLUXNET database. For example, there were 39 sites in the temperate forest biome but only three in the boreal forest biome. ML models were trained for every available biome, but under-represented biomes are less likely to generalize to new locations until data from additional sites of that biome can be leveraged for model training. The mean and standard deviation of the input features and sap flux for each cluster were collected to investigate model differences. Box plots summarizing the mean, distribution, and spread of the data are presented in Supplementary Information B. 

\subsection{Machine learning algorithm configuration}

Two supervised-based ML models (neural networks and random forests) were developed for transpiration prediction. A grid search based on k-fold cross-validation approach with k=5 \citep{Pedregosa2011} was implemented to select the best hyperparameters for each model. Cross-validation searches split the data into equal parts, hold out one part during training, and test the model's performance on the excluded data. Model hyperparameters describe the architecture of the model. For neural networks, the hyperparameters include the number of hidden layers, the number of neurons in each hidden layer and the number of epochs for training. The hidden layers are the layers of neurons between the input and output layers. Each epoch represents one pass through the training data. The hyperparameters of random forest models are the number of estimators and the maximum depth of each estimator. The number of estimators refers to the number of decision trees in the random forest. The maximum depth is the number of branches in each decision tree. Limiting the maximum depth of a tree will favor simpler models, which can prevent overfitting. The ranges of all hyperparameters used in the grid search are summarized in Supplementary Information C.

The fixed hyperparameters (learning rate and mini-batch size) were necessary to constrain the neural network architecture so that the model selection algorithm could be performed in a reasonable time frame. The neural network algorithm used the mean squared error as the loss function for training.  The learning rate was fixed at 0.001. The neural network training algorithm used an adaptive moment estimation (Adam) optimizer, which incorporates the concept of momentum to accelerate the gradient descent algorithm and use less memory \citep{chollet2021deep}. The mini-batch hyperparameter was set to 32, so each mini-batch contained 32 data points. The mini-batch hyperparameter was used to speed processing and reduce the weight on any individual data point. The activation function used in the input and hidden layers was the rectified linear unit (ReLU).

\subsection{Analysis of model performance and feature importance}

The $R^2$ fitness and mean absolute error ($MAE$) \citep{McClave2006} for each set of model predictions were computed in reference to the ground-truth data as the primary measure of model performance, using the appropriate sci-kit learn metrics packages \citep{Pedregosa2011}. Two techniques, calculation of the Spearman correlation coefficient and permutation-based feature importance tests, were used to analyze the importance of different input features (i.e., environmental variables) to the target value (i.e., sap flow as a surrogate to transpiration rate). The Spearman correlation coefficient \citep{Virtanen2020} was calculated for each cluster (i.e., PFT and biome) to measure the strength of the monotonic relationship between each input feature and target value. Permutation-based feature importance tests \citep{Pedregosa2011} were performed for all models to determine the sensitivity of the modeled transpiration to each input feature, measuring the relative decrease in model performance score when the data in the testing set related to an input feature is randomly shuffled among the testing samples. The test was implemented using the scikit-learn package in Python \citep{Pedregosa2011}. The permutation-based feature importance can be compared to the monotonic strength of the relationship in the data (i.e., the Spearman correlation coefficient) to better understand the relationship underlying the feature importance. For example, if there is no monotonic relationship but the model feature importance is high in a high-performing model, there is likely a non-monotonic relationship that was not captured by the Spearman correlation coefficient. 

\section{Results}

\subsection{Model performance}

The best random forest and neural network models were selected using grid search and cross-validation, with resulting model hyperparameters reported in Table C.2. In general, the grid search selected model hyperparameters that indicated greater model complexity: more epochs, hidden layers, and neurons in the neural network models, and a greater maximum depth, and number of estimators in the random forest models. 

Model performance is characterized by the cross-validation score, i.e., the average validation score in terms of $R^2$ across all five cross-validation folds. The model performance varies considerably between different clusters and ML models, as reported in Table D.1. Summarized based on clustering strategies, Figure \ref{fig:r2_feat_imp}a shows the variation of model performance associated with different strategies. It indicates that, on average, clustering based on biome results in improved model performance over clustering by PFT, although the deciduous and mixed clusters scored significantly higher in both the neural network and random forest models than the evergreen cluster. It can also be seen that the random forest models in general perform better than the neural network models.

\subsection{Feature importance}

For each cluster, the Spearman correlation coefficient, which indicates the strength of the monotonic relationship between the target and the input variable, was calculated between sap flux and each environmental variable (see Figure \ref{fig:transpiration_correlation}), with biome types arranged in order of descending MAP. Most input variables exhibited positive monotonic relationships with sap flux, with the exception of PPFD in the subtropical desert cluster, and SWC in the tropical rainforest and woodland/shrubland clusters. The median and spread of the feature importance data for each algorithm overall is shown in Figure \ref{fig:r2_feat_imp}b. For each data cluster, individual relative model feature importances are presented in Figure \ref{fig:feat-importances}. Interestingly, there was high variability in the average feature importance between models. In the random forest models, PPFD was, on average, the most important input feature, followed closely by SWC. Air temperature and VPD were somewhat less important. In the neural network models, however, air temperature was the most important feature, followed by PPFD, VPD, and then SWC. 

\section{Discussion}

The use of a global sap flux dataset enabled the development of ML algorithms trained on 95 sites around the world which are clustered across a number of different sites, biomes, and plant species to provide insights into global drivers of forest transpiration. The model performance results indicated that ML can be used successfully to train general models of sap flux provided that data is appropriately clustered, while the significant differences in feature importance between high-performing models showed that the key predictors of transpiration vary by biome and PFT.

\subsection{Trends in model performance}

\begin{figure}
    \centering
    \includegraphics[width=12cm]{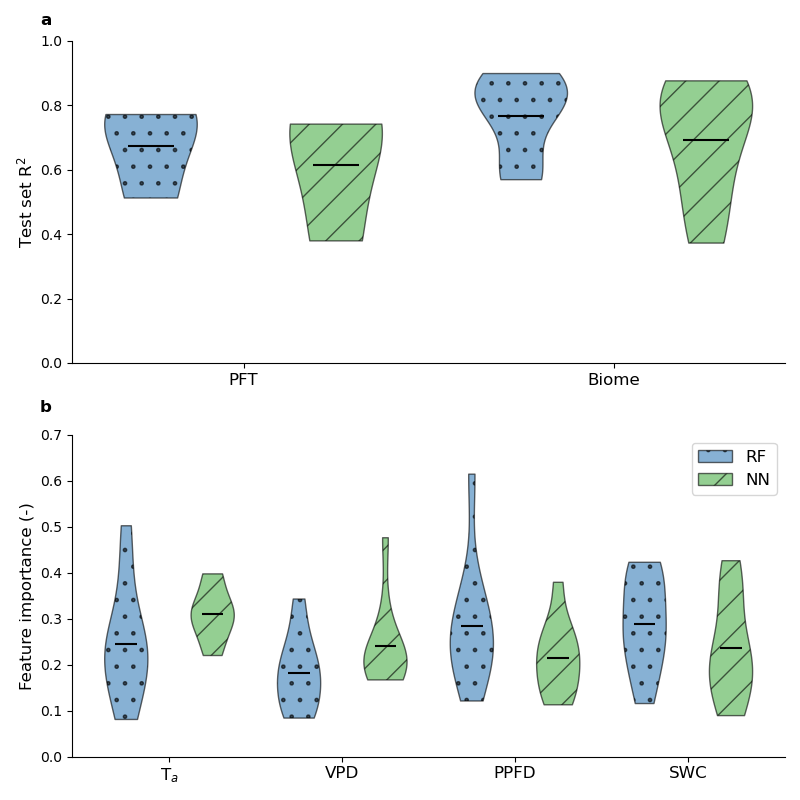}
    \caption{(a) Mean and spread of test set $R^2$ statistics of the random forest (blue, dotted) and neural network (green, hatched) models for each clustering strategy (plant functional type and biome). (b) Mean and spread of relative feature importances for air temperature ($T_a$), vapor pressure deficit (VPD), photosynthetically active photon flux density (PPFD), and shallow soil water content (SWC) using the random forest (RF) and neural network (NN) algorithms.}
    \label{fig:r2_feat_imp}
\end{figure}

The model performance of individual clusters varied widely, with $R^2$ values ranging from 0.37 to 0.90 (Figure \ref{fig:r2_feat_imp}a and Supplementary Information D). The boreal forest and temperate grassland desert biomes each had three co-located sites and achieved $R^2$ values similar to the broader literature (0.86 and 0.83, respectively). For studies that use data from sap flow sensors installed in trees at a single study site, it is not uncommon to achieve $R^2$ values of greater than 0.9 \citep{Liu2009, Tu2019, Ellsaber2020}. In such studies, the species and climate do not vary, so the models achieve higher accuracy but are unlikely to provide generalizable insights into transpiration drivers. As individual clusters with high numbers of sites, the deciduous cluster and the woodland/shrubland biome cluster produced models with good performance for 35 and 36 different sites around the globe, respectively. These clusters contained varying species and climates, yet achieved $R^2$ values of 0.77 in the deciduous cluster and 0.80 in the woodland/shrubland biome cluster. While the reported performance on the data set is not as high as for models trained on a single location, the performance is still relatively high and, importantly, the models will reflect more robust and generalizable trends in ecosystem response to climate factors. The performance in some clusters surpassed the performance of models trained on data clusters from \cite{Ferreira2019}, although their target was reference ET rather than measured transpiration. Further tuning of hyperparameters and incorporation of the effect of time lag between changes in environmental variables and sap flux response \citep{Wang2017, Tu2019} could improve the accuracy of these models.

\begin{figure}
    \centering
    \includegraphics[width=15cm]{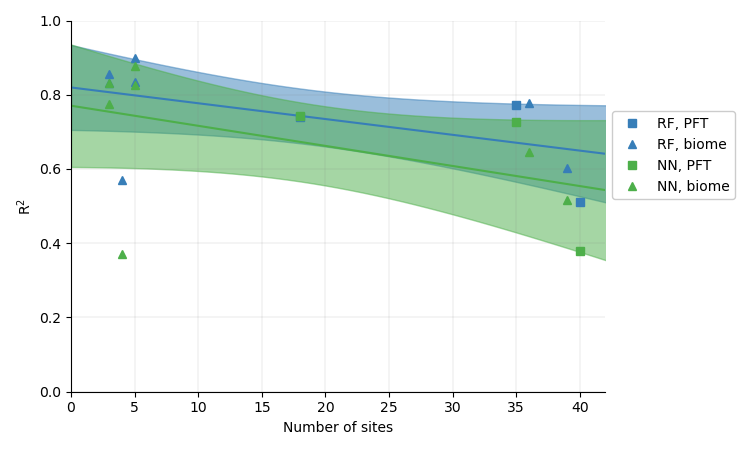}
    \caption{A weak negative correlation was observed between the number of sites included in each model and the $R^2$ model fitness for both the random forest (RF) and neural network (NN) models.}
    \label{fig:performance-site-correlation}
\end{figure}

The relationship between the number of sites and model fitness displayed in Figure \ref{fig:performance-site-correlation} is, at first glance, counter to the typically observed trend, in which more data produces a better ML model. In this case, a cluster that is too large does not have a common underlying pattern between sites to predict transpiration. The best-performing clusters were large enough to act as a general model, but not so general that controls on transpiration differed significantly within the cluster. While the performance metrics were highest for small clusters, mid-sized clusters will be more useful to researchers searching for a general model to fit to a new location, because the smallest clusters contained too few sites to be expected to generalize to new locations. Mid-sized clusters are also more representative of the controls on transpiration of that cluster, since the addition of sites reduces the likelihood that underlying site-specific patterns are controlling the model. Models trained on clusters that were too large lost meaning and in some cases may be less useful because they failed to predict transpiration with a reasonable degree of accuracy. This trend in performance shows that sensitivity of transpiration to meteorological variables does significantly differ between sites, but that some sites share enough similarities to fit into the same general model. The cross-validation search and limitations on model complexity made overfitting unlikely. The correlation between the number of sites and the resulting model fitness is shown using a simple linear regression model in Figure \ref{fig:performance-site-correlation}. There was a weak negative correlation ($R^2$ = 0.29 for RF model) between the number of sites and the $R^2$ model fitness. Some groups with a small number of sites may be overfit to site-specific conditions which are not necessarily captured in the model. This likely applies to the boreal forest and temperate grassland desert biomes, which only contain co-located sites. Other models, such as the tropical rainforest and tropical forest savanna biomes, had a small number of sites, but were not co-located, decreasing the likelihood that the model was overfit to variables not captured by the model.

Between the two tested clustering strategies, the biome clusters had the highest median performance of 0.83, as compared with a median performance of 0.73 for the PFT clusters (Figure \ref{fig:r2_feat_imp}). The greater performance of the biome clusters may be primarily due to the smaller average sample size of these clusters (see Figure \ref{fig:performance-site-correlation}). While the deciduous cluster significantly outperformed the predicted model performance based on its number of included sites, its performance was very similar to that of the woodland/shrubland cluster with a similar number of sites (R$^2$ value of 0.77 and 35 sites as compared with an R$^2$ value of 0.78 and 36 sites). At the same time, the evergreen cluster with 40 sites significantly underperformed relative to the average trend, while the mixed cluster had an R$^2$ value of 0.74, very close to that expected based on its 18 sites. Within the functional type clustering strategy, the deciduous and mixed clusters had significantly higher performance than the evergreen cluster. The distribution of sap flux and recorded environmental variables for this cluster did not differ significantly from that of the other clusters, so it is likely that the decreased performance on evergreen trees could indicate that evergreen trees are more dependent on environmental or physiological variables that were not included in this study, that the data on evergreen trees was lower quality than the data included in other clusters, or that the evergreen functional type had a wider distribution of plant functional traits, causing difficulty in predicting overall evergreen responses to the environment. Finally, the evergreen cluster may have been too large, meaning that the model was unable to determine the underlying patterns because they were different between sites in the cluster. The evergreen cluster had the greatest amount of data: 40 sites and 1,602,286 data points, as compared with the deciduous cluster, which had the second-highest amount of data with 35 locations and 254,402 data points. All of this suggests that the evergreen cluster may need further subdivision for creation of generalizable models. Investigation of these possibilities could reveal new understanding of the controllign factors of evergreen forest transpiration. 

The random forest models had overall higher performance than the neural network models (Figure \ref{fig:r2_feat_imp}a), with an average R$^2$ value of 0.74 compared with 0.67. It is possible that transpiration prediction is a problem that lends itself well to the random forest model architecture. Random forests are built by a collection of individual estimators (decision trees) that rely on bifurcating splits, such as the difference between daytime and nighttime transpiration. In many models, nighttime transpiration is near-zero because the tree does not need to photosynthesize and closes its stomata as a response to zero incoming radiation. Since its architecture specializes on detecting binary decisions, the random forest algorithm could be expected to excel at detecting this difference, and possibly other particulars of transpiration prediction. The second possibility is that an aspect of the artificial neural network architecture is deficient. The possible permutations of hyperparameters used to create neural networks are limitless, and a limited number of neural network architectures can be tested. There is a vast literature dedicated to problem-specific design and implementation of neural networks, and multiple papers within the transpiration prediction literature have tackled this problem \citep{Tu2019, Yan2021}. Comparing average performance results between the neural network and random forest models shows that the neural network models built and tested in this study, although outperformed by the random forest models, were not far behind. Future studies could test different permutations of neural network architecture, perhaps following the guidance of \cite{Lakshmanan2020}.

\subsection{Trends in climate and feature importance}

\begin{figure}
    \centering
    \includegraphics[width=12cm]{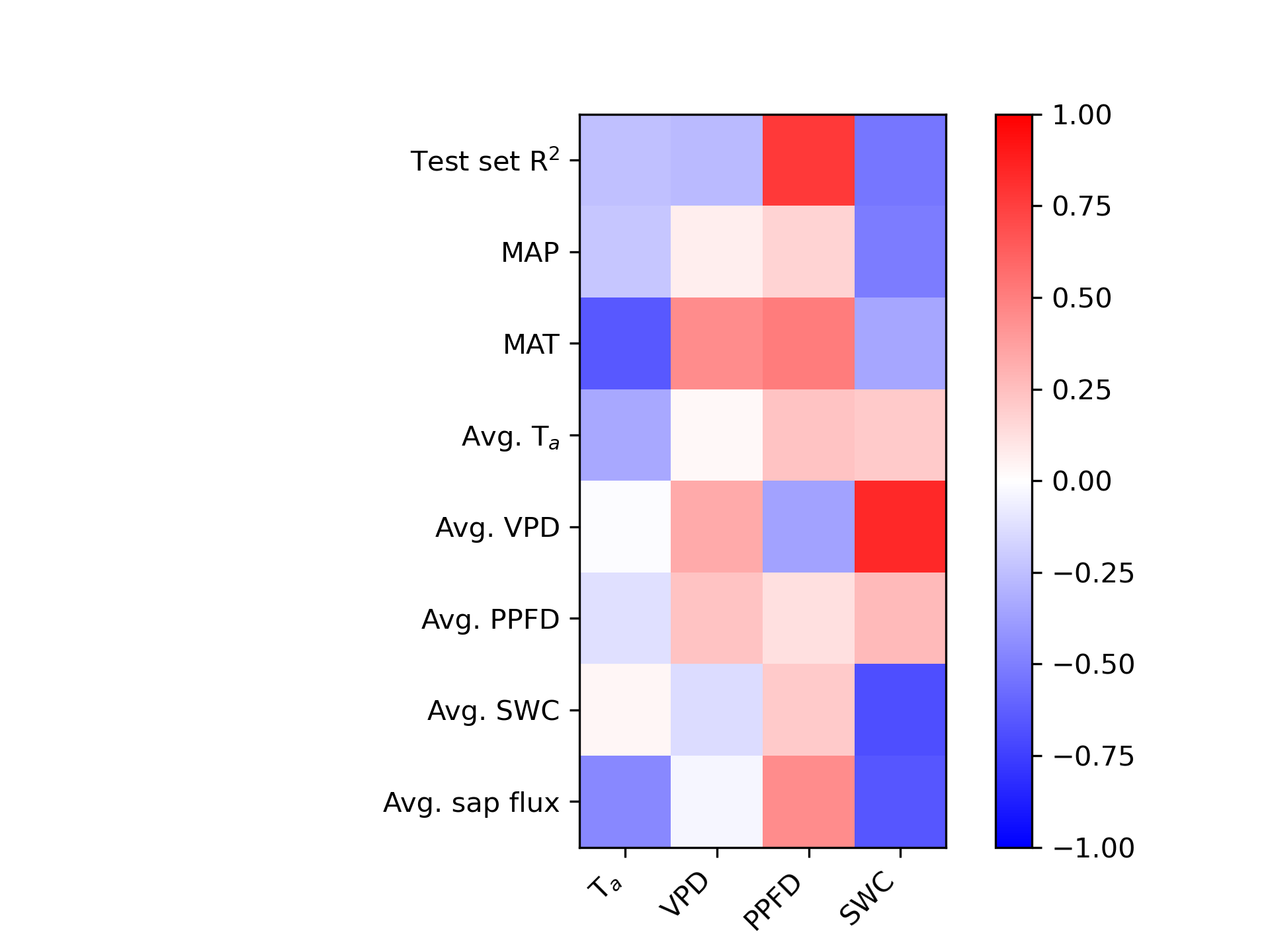}
    \caption{Spearman correlation between feature importances of input variables (air temperature ($T_a$), vapor pressure deficit (VPD), photosynthetically active photon fluid density (PPFD) and shallow soil water content (SWC)) and climate statistics (mean annual precipitation (MAP), mean annual temperature (MAT), average air temperature (Avg.$T_a$), average photosynthetically active photon flux density (Avg. PPFD), average shallow soil water content (Avg. SWC) and average sap flux (Avg. sap flux).}
    \label{fig:climate_correlation}
\end{figure}

\begin{figure}
    \centering
    \includegraphics[width=15cm]{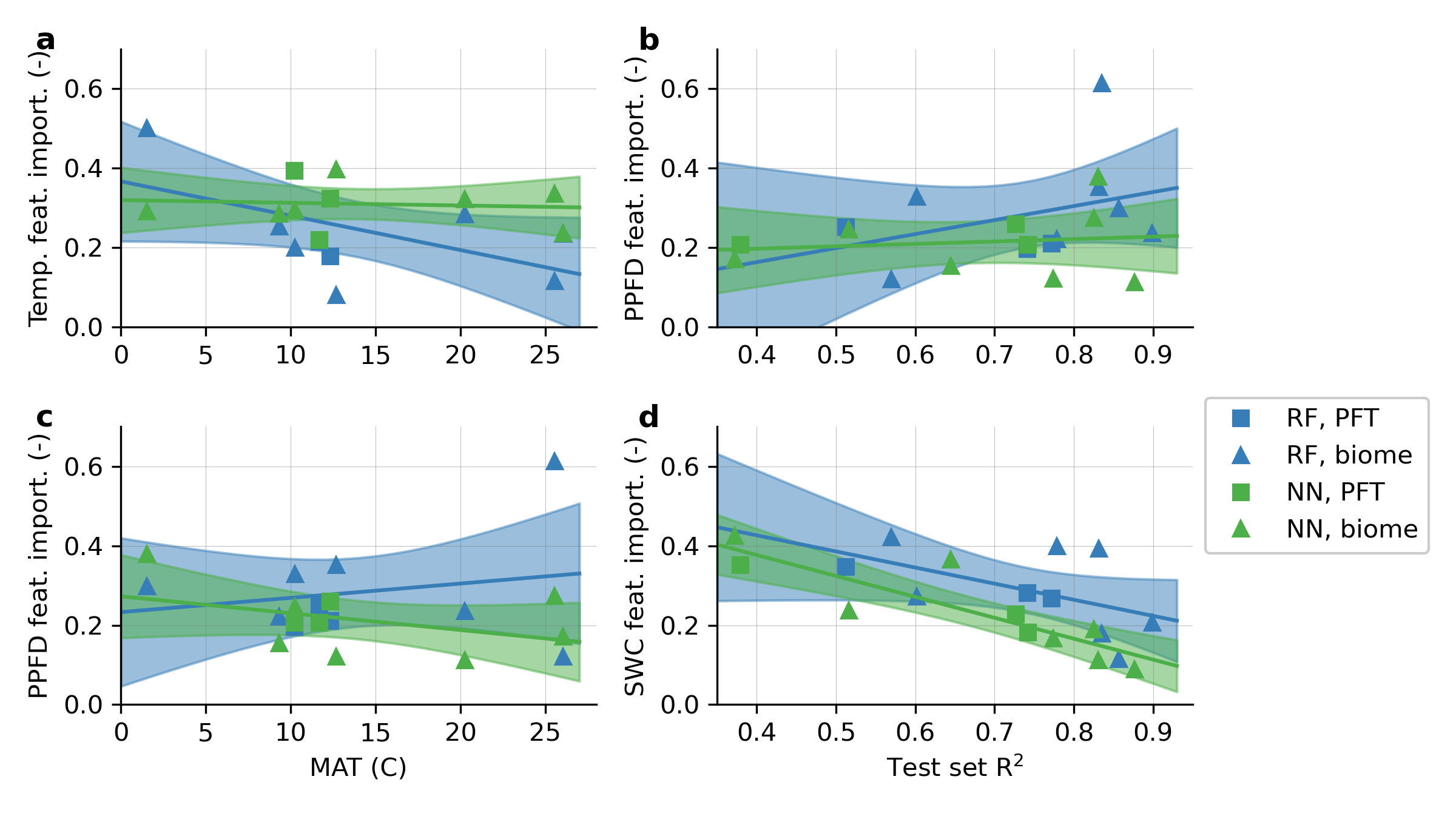}
    \caption{Correlations observed between input feature importances (air temperature,  photosynthetically active photon fluid density (PPFD), and shallow soil water content (SWC)) and the mean annual temperature (MAT) (a, c) and test set $R^2$ (b, d) of the cluster for random forest (RF, blue) and neural network (NN, green) models.}
    \label{fig:mat-r2-panels}
\end{figure}

\begin{figure}
    \centering
    \includegraphics[width=15cm]{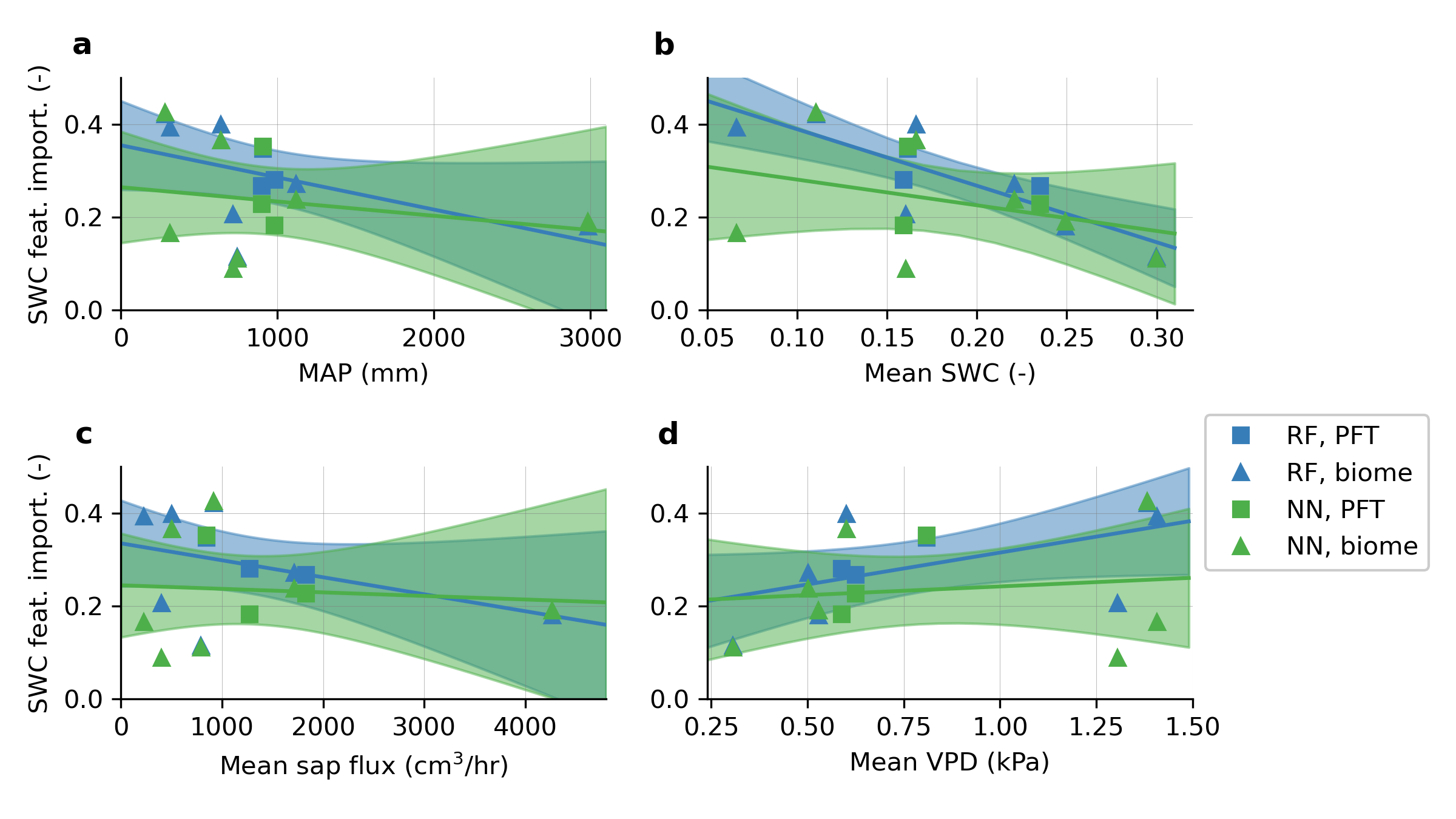}
    \caption{Correlations exhibited between the soil water content (SWC) feature importance of the cluster and average climate conditions including mean annual precipitation (MAP), mean shallow soil water content (SWC), mean sap flux, and mean vapor pressure deficit (VPD) for random forest (RF, blue) and neural network (NN, green) models.}
    \label{fig:swc-panels}
\end{figure}

Variation of feature importance by climate was analyzed to find trends between climate statistics and relative feature importance of meteorological variables. Results from the random forest models are primarily discussed here because they out-performed the neural network models for each cluster. Correlations were examined between the feature importance of each environmental variable and the mean annual temperature (MAT), mean annual precipitation (MAP), average statistics for every feature, and the test set $R^2$. The correlation matrix is shown in Figure \ref{fig:climate_correlation}. For pairings with the strongest monotonic relationships (defined as a Spearman correlation coefficient with an absolute value greater than 0.5), the results from the random forest algorithm were plotted in Figure \ref{fig:mat-r2-panels} and Figure \ref{fig:swc-panels} to illustrate the relationship between climate statistics and the feature importance of various input features; these results are discussed in the subsequent sections. The feature importance of air temperature was found to have a weak negative correlation with the MAT, as shown in Figure \ref{fig:mat-r2-panels}a. The feature importance of VPD was not found to have a strong monotonic relationship with model performance or with any of the climate statistics shown. However, the average VPD had an influence on the feature importance of SWC (Figure \ref{fig:swc-panels}d).

The solar radiation model feature importance had a strong monotonic relationship with the model performance, and a moderately strong monotonic relationship with the mean annual temperature (Figure \ref{fig:mat-r2-panels}b and c). The shallow SWC feature importance exhibited a monotonic relationship with most of the modeled climate statistics, as well as the model performance, indicating that transpiration sensitivity to SWC is highly dependent on the climate and may be predictable based on climate type. The shallow SWC feature importance exhibited a negative correlation with MAP, mean shallow SWC, the test set $R^2$ measuring model performance, and the average sap flux (Figure \ref{fig:mat-r2-panels}d and \ref{fig:swc-panels}). Mean annual VPD exhibited a positive correlation with the importance of SWC in the model.

\subsection{Insights into drivers of transpiration}

\begin{figure}
    \centering
    \includegraphics[width=12cm]{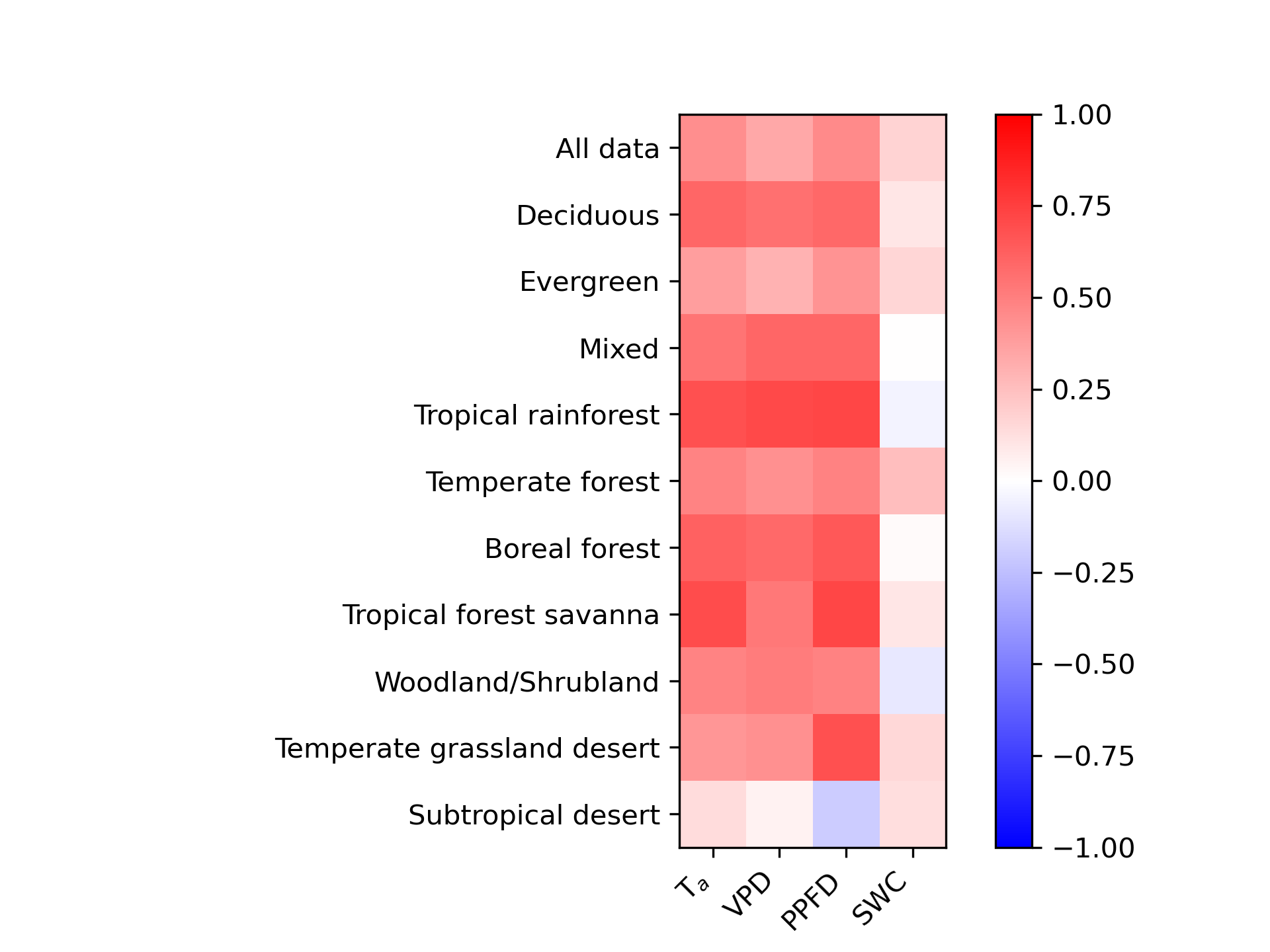}
    \caption{Spearman correlation between each environmental variable (air temperature ($T_a$), vapor pressure deficit (VPD), photosynthetically active photon flux density (PPFD), and shallow soil water content (SWC)) and sap flux for every data cluster, with biome types arranged in order of descending mean annual precipitation.}
    \label{fig:transpiration_correlation}
\end{figure}

Feature importance and calculation of the Spearman correlation coefficient showed that the primary variables controlling transpiration varied significantly between clusters (Figure \ref{fig:transpiration_correlation}). In general, air temperature, VPD, and PPFD all exhibited a strong positive monotonic relationship with transpiration, except in the case of the subtropical desert cluster, which supports other findings that high atmospheric water demand can limit transpiration under water-limited conditions \citep{Sulman2016, liu2020plant}. Previous studies investigating feature importance for sap flux prediction have mainly focused on single locations. \cite{Liu2009} studied sap flow in pear trees in a temperate climate and found that VPD, PPFD, air temperature, wind speed, SWC, and leaf area index data all improved sap flow prediction, but that VPD and PPFD were the most important features. \cite{Tu2019} studied a monoculture plantation of pine trees with high average precipitation and temperature and found that RH and VPD were the most important predictors of sap flow. \cite{Shiri2014} investigated multiple locations and demonstrated that ML models generally had better performance in more humid regions than more arid regions. 

The relationship between SWC and transpiration was also expected to be positive, since water availability is critical for transpiration. Instead, a weak correlation, and sometimes a weak negative correlation, was observed between SWC and transpiration. It is well-known that SWC is non-linearly related to transpiration, with minimal to no response of transpiration to SWC after SWC surpasses the water stress point \citep{porporato2002ecohydrology}. It is possible that SWC varies on a longer time scale than the other parameters, and thus is interpreted by ML models as a proxy for another variable, such as time of year, which is not represented in the data; that the SWC data may be of lower quality than the other inputs; or that the shallow SWC measured in the SAPFLUXNET studies does not accurately represent total water availability in the system. The weak negative relationship could also be due to lower solar radiation during precipitation events that are associated with increases in SWC. Despite the absence of a strong positive monotonic relationship in the data, SWC was still an important feature for many of the ML models. This means that either a non-monotonic relationship is present, or SWC was used by the model as an proxy of an unrecorded independent variable. There is some support for the first hypothesis, as \cite{Fu2022} found that gross primary production, which is correlated with transpiration, increases in response to decreasing SWC when SWC is high, and the relationship is only positively monotonic when SWC is below a certain threshold.

\begin{figure}
    \centering
    \includegraphics[width=16cm]{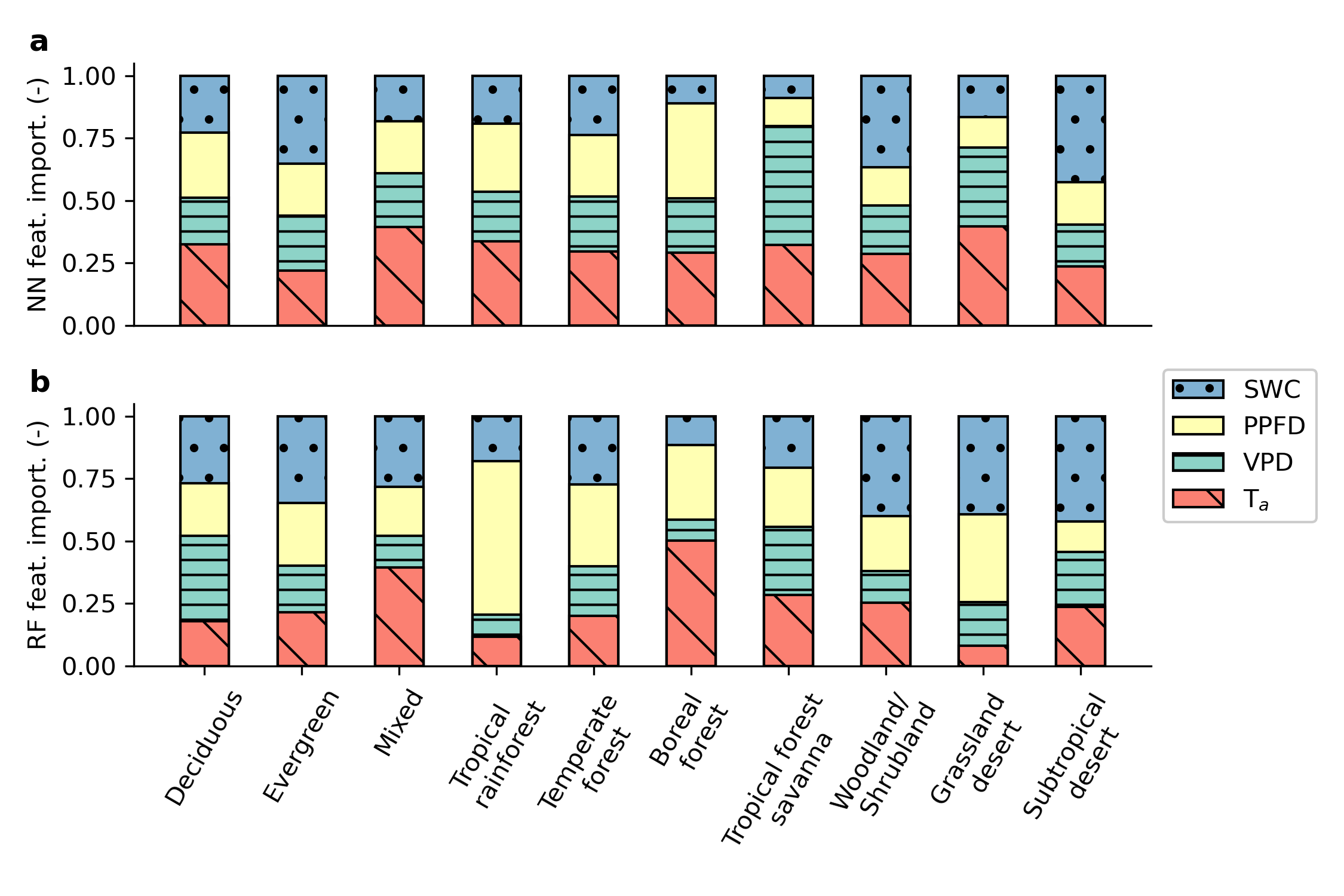}
    \caption{Relative feature importances of air temperature ($T_a$), vapor pressure deficit (VPD), photosynthetically active photon flux density (PPFD) and shallow soil water content (SWC) for each (a) neural network (NN) and (b) random forest (RF) model, with biome types arranged in order of descending mean annual precipitation.}
    \label{fig:feat-importances}
\end{figure}

While the Spearman correlation was useful for examining data quality and whether the relationship was monotonic, a higher Spearman correlation coefficient between a meteorological variable and the measured sap flow did not always predict a higher feature importance for that variable in the model. This was true not only for SWC, but for the other environmental variables as well (Figures \ref{fig:transpiration_correlation} and \ref{fig:feat-importances}). The strength of the monotonic relationship in the data could not be used to predict which variables would control transpiration in the model, even for high-performing models, showing that ML is a uniquely useful tool for this problem because there are complex, nonlinear interactions among variables that predict transpiration. 

The findings of this study indicate that the relative importances of SWC and VPD vary by climate, which may complicate the ongoing debate over their respective importances for governing transpiration \citep{Novick2016, Sulman2016, liu2020plant, Liu2020}. Solar radiation and VPD were important to every model (Figure \ref{fig:r2_feat_imp}b), but their feature importances exhibited high variance between clusters ( Figure \ref{fig:feat-importances}). A growing literature has highlighted the importance of VPD. According to the findings of \cite{Novick2016}, the importance of VPD will increase as the climate changes due to near-universal increases in VPD, and despite being a key predictor of transpiration in many systems, it is often overlooked as a factor in ecohydrological models. These results highlight the importance of current ongoing efforts to include plant hydraulics in land surface models \citep{Kennedy2019}.

As exhibited in Figure \ref{fig:swc-panels}, weak correlations were detected between the importance of SWC and various measures of water availability, indicating that SWC may be a more important input feature in water-limited climates, such as the temperate grassland desert and woodland/shrubland biomes in this study. As water becomes scarce, transpiration is increasingly dependent on the availability of water. At high water availability, transpiration always occurs but will reach a plateau, and this will occur more quickly if there is a light or VPD limitation. These findings are corroborated by the findings of \cite{Novick2016}, who found that SWC limitations to transpiration tended to increase monotonically with dryness index. Their finding that the importance of VPD is highly variable between climates is also supported. The relationship between water availability and the importance of SWC is also consistent with \cite{Sulman2016}, who observed that SWC had the largest influence on transpiration when SWC was low.  \cite{Liu2020} found that after decoupling VPD and SWC, SWC was most important in semi-arid regions such as shrubland, grassland and savanna, and VPD effects are weakest in these ecosystems. Their finding that the importance of SWC is dominant over VPD was not corroborated. The contrary findings of \cite{Sulman2016} that VPD is at least equally important as SWC, and often dominating, was supported. A possible explanation for this is that \cite{Liu2020} examined variable importance at an annual scale, whereas \cite{Sulman2016} investigated importance at a finer temporal scale, noting that VPD changes more rapidly than SWC, so VPD would be expected to be more important at hourly time scales.

\section{Conclusion}

This work analyzed the results of two ML methods (neural networks and random forests) for creating clustered models to predict forest transpiration across a wide range of biomes and tree species. The results suggest that creating clusters for ML models is an advisable approach for future transpiration modeling, as many models achieved good results for clusters as large as 35-36 sites. Smaller clusters with up to three co-located sites achieved R$^2$ values similar to the broader literature for models at a single study site. While inclusion of a greater number of sites lowered model performance in general, the models trained on a wide variety of sites are more likely to generalize to novel locations. Thus, a tradeoff exists between model accuracy and generalizability because models trained on few or co-located sites tend to reflect site-specific conditions. A site-specific model may be more useful for some applications, but a generalizable model could be very useful for researchers attempting to estimate transpiration rates in locations with little to no field data.

The clustering of input data by biome and PFT suggestes that one method is not clearly preferable to the other, rather, it seemed that the strongest control on model accuracy was the total number of sites included in the model. Because this study only included trees, the considered PFTs were limited to evergreen and deciduous trees. It is expected that, if more diverse PFTs such as grasses or shrubs were considered, PFT might exert a stronger control on the model results. In future work, it would be interesting to consider other methods for data cluster creation for transpiration modeling, including generation of groups using a k-means clustering algorithm which could consider many factors simultaneously.

Analysis of Spearman correlation coefficients between input variables and sap flux as well as feature importance of input variables in the ML models indicated that drivers of transpiration varied strongly across biome type and ML method. The random forest results, which performed slightly better than the neural network results, suggested that, while $T_a$, VPD, PPFD, and SWC all contributed significantly to model performance, PPFD and SWC made the largest contributions. The relationships between the feature importances of these two variables exhibited a loose negative correlation, suggesting that SWC tended to the the most important predictor in the biomes with the lowest MAP, while PPFD was more important in biomes with higher MAP. Results also highlighted the importance of VPD in making predictions of transpiration, suggesting that detailed plant hydraulics models which fully account for the impacts of atmospheric dryness on plant water stress may contribute to better future predictions of mechanistic vegetation models. 

\section*{CRediT authorship contribution statement}

\textbf{Morgan Thornwell:} Conceptualization, Methodology, Investigation, Writing - original draft. \textbf{David Yang:} Conceptualization, Methodology, Writing - review and editing. \textbf{Cheng-Wei Huang:} Writing - review and editing. \textbf{Peyman Abbaszadeh:}Writing - review and editing.
\textbf{Samantha Hartzell:} Conceptualization, Methodology, Supervision, Writing - original draft, Writing - review and editing, Resources.

\section*{Acknowledgments}

This work was supported by the Department of Energy (DOE) through the Office of Biological and Environmental Research (BER) (DE-SC0023468) and the National Science Foundation (NSF-CBET-2139003, NSF-ORE-CZ-2423295). We thank Roberto Palacios for his useful feedback.

\section*{Data availability}
Data used in the analysis is publicly available in the SAPFLUXNET database \citep{poyatosGlobalTranspirationData2021}. The models built in this study, as well as the code used to train them, are publicly available on Github at https://github.com/mortholl/Transpiration-ML-Project for open source use.

\clearpage

\clearpage

\end{document}